\documentclass[showpacs,preprintnumbers,amsmath,amssymb]{revtex4}
\usepackage{graphicx}
\begin{document}

\date{\today}
\vspace{2.7in}

\title{Bose-Einstein condensation of trapped polaritons in 2D electron-hole systems in a high magnetic field}
\author{Oleg L. Berman$^{1}$, Roman Ya. Kezerashvili$^{1,2}$, and Yurii E.
Lozovik$^{3}$} \affiliation{\mbox{$^{1}$Physics Department, New York
City College of Technology, The
City University of New York,} \\
Brooklyn, NY 11201, USA \\
\mbox{$^{2}$The Graduate School and University Center, The
City University of New York,} \\
New York, NY 10016, USA \\
\mbox{$^{3}$Institute of Spectroscopy, Russian Academy of
Sciences,} \\
142190 Troitsk, Moscow Region, Russia}

\begin{abstract}

The Bose-Einstein condensation (BEC) of magnetoexcitonic polaritons
in two-dimensional (2D) electron-hole system embedded in a semiconductor microcavity in a high
magnetic field $B$ is predicted. There are two physical realizations of 2D electron-hole system under consideration: a graphene layer and quantum well (QW). A 2D gas of magnetoexcitonic polaritons
is considered in a planar harmonic potential trap. Two possible
physical realizations of this trapping potential are assumed:
inhomogeneous local stress or harmonic electric field potential  applied to excitons and a parabolic
shape of the semiconductor cavity causing the trapping of
microcavity photons. The effective Hamiltonian of the ideal gas of
cavity polaritons in a QW and graphene in a high magnetic field and the BEC
temperature as functions of magnetic field are obtained. It is
shown that the effective polariton mass $M_{\rm eff}$ increases with
 magnetic field as $B^{1/2}$. The BEC
critical temperature $T_{c}^{(0)}$ decreases as $B^{-1/4}$ and
increases with  the spring constant of the parabolic
trap. The Rabi splitting related to the creation of a magnetoexciton in a high magnetic field in graphene and QW is obtained. It is shown that Rabi splitting in graphene can be controlled by the external magnetic field  since it is proportional to $B^{-1/4}$, while in a QW the Rabi splitting does not depend on the magnetic field when it is strong.

\vspace{0.1cm}



\end{abstract}

\maketitle {}


\section{Introduction}
\label{intro}

In the past decade,  Bose coherent effects of 2D excitonic
polaritons in a quantum well embedded in a semiconductor microcavity have been the subject of theoretical and
experimental studies  \cite{pssb,book}. To obtain polaritons, two mirrors placed opposite each other form a microcavity, and quantum wells are embedded within the cavity at the antinodes of the confined
optical mode. The resonant exciton-photon interaction results in the Rabi splitting of the excitation spectrum.  Two polariton branches appear in the spectrum due to the resonant exciton-photon coupling. The lower polariton (LP) branch of the spectrum has a minimum at zero momentum. The effective mass of the lower polariton is extremely  small, and lies in the range $10^{-5}-10^{-4}$ of the free electron mass. These lower polaritons form a 2D weakly interacting Bose gas. The extremely light mass of these bosonic quasiparticles, which corresponds to
experimentally achievable excitonic  densities, result in a relatively high
critical  temperature for superfluidity, of $100 \
\mathrm{K}$ or even higher. The reason for such a high  critical temperature is that the 2D thermal de Broglie wavelength is inversely proportional to the mass of the quasiparticle.

While at finite temperatures there is no true BEC  in any infinite untrapped
2D system, a true 2D BEC quantum phase transition can be obtained in the presence of a confining
potential \cite{Bagnato,Nozieres}.
  Recently,  the  polaritons in a harmonic potential trap have  been studied experimentally in a GaAs/AlAs quantum well embedded in a GaAs/AlGaAs microcavity \cite{Balili}. In this trap,
   the exciton energy is shifted using a stress-induced
band-gap. In this system, evidence for the BEC of polaritons in a quantum well
has been observed  \cite{science}. The theory of the
BEC and superfluidity of excitonic polaritons in a quantum well
without magnetic field in a parabolic trap has been developed in
Ref.~[\onlinecite{Berman_L_S}]. The Bose condensation of polaritons is caused by their
bosonic character \cite{science,Berman_L_S,Kasprzak}.

While the 2D electron system was studied in quantum wells \cite{Ando_Fowler_Stern} in the past decade, a novel type of 2D electron system was experimentally obtained in  graphene,
 which is a 2D honeycomb lattice of the carbon atoms that form the basic planar structure in graphite
 \cite{Novoselov1,Zhang1}.
Due to unusual properties of the  band
structure, electronic properties of graphene became the object of many recent experimental and
theoretical studies \cite{Novoselov1,Zhang1, Novoselov2,Zhang2,Falko,Katsnelson,Castro_Neto_rmp}.  Graphene is a gapless
semiconductor with massless electrons and holes which have been
described as Dirac-fermions \cite{DasSarma}. The unique electronic properties  in graphene in a magnetic field have been studied recently \cite{Nomura,Jain,Gusynin1,Gusynin2}. The  electron-photon
interaction in graphene was discussed, for example, in Ref.~[\onlinecite{Vafek}].
The energy spectrum and the wavefunctions of magnetoexcitons, or electron-hole pairs in a magnetic field, in graphene have been calculated in interesting works~\cite{Iyengar,Koinov}.

 The  spatially-indirect excitons  in coupled quantum wells (CQWs), with and without  a magnetic field $B$ have been studied recently experimentally in Refs.~[\onlinecite{Snoke,Butov,Timofeev,Eisenstein}].
 The experimental and theoretical interest in these systems is  particularly due to the possibility of the BEC
and superfluidity of indirect excitons, which
can manifest  in the CQW as persistent electrical currents
in each well and also through coherent optical properties and
Josephson phenomena \cite{Lozovik,Littlewood,Vignale,Berman}.
Since the exciton binding energies increase with magnetic field,  2D magnetoexcitons survive in a substantially wider temperature range in high magnetic fields \cite{Lerner,Paquet,Kallin,Yoshioka,Ruvinsky,Ulloa,Moskalenko}.
The BEC and superfluidity of spatially-indirect magnetoexcitons with
spatially separated electrons and holes have been studied in
graphene bilayer \cite{Berman_L_G} and graphene superlattice
\cite{Berman_K_L,Berman_K_L_2}. The electron-hole pair condensation in the
graphene-based bilayers have been studied in~[\onlinecite{Sokolik,MacDonald1,MacDonald2,Efetov}]. However, the polaritons in graphene in high magnetic field have not yet been considered. The BEC and superfluidity of cavity polaritons in a QW without a trap were considered in~[\onlinecite{Kavokin2,Kavokin3}]. It is interesting to study a 2D system such as polaritons in graphene embedded in a microcavity from the point of
view of the existence of the BEC within it.

The purpose of this paper is to point out the existence of the BEC
of the magnetoexcitonic polaritons in a QW and a graphene layer embedded in a
semiconductor microcavity in a strong magnetic field and to discuss the
condition of its realization. Since it was shown that the magnetoexcitons in a QW and graphene layer in a high magnetic field can be described by the same effective Hamiltonian with the different effective mass of a magnetoexciton \cite{Berman_K_L,Berman_K_L_2}, we expect to obtain the similar expressions for the critical temperature of BEC for cavity polaritons  in a QW and graphene with the only difference in the effective mass of a magnetoexciton.

The paper is organized in the
following way. In Sec.~\ref{is_mag}  the spectrum of an isolated
magnetoexciton with the electron and hole in a single graphene layern and QW
is derived by applying  perturbation theory with respect to the strength of the Coulomb
electron-hole attraction. In Sec.~\ref{Hamil_eff} the effective
Hamiltonian of microcavity polaritons in graphene and QW in a high magnetic
field along with a trapping potential is derived. In Sec.~\ref{Rabi} the Rabi splitting related to the creation of a magnetoexciton in graphene and QW in a high magnetic field is obtained. The
temperature of BEC and the number of polaritons in Bose-Einstein
condensate as a function of temperature, magnetic field and spring
constant are calculated in Sec.~\ref{bec}. Finally, the discussion of the results and
 conclusions follow in Sec.~\ref{discussion}.


\section{An isolated magnetoexciton in a single graphene layer and QW}
\label{is_mag}

\begin{figure}[t] 
   \centering
  \includegraphics[width=3.5in]{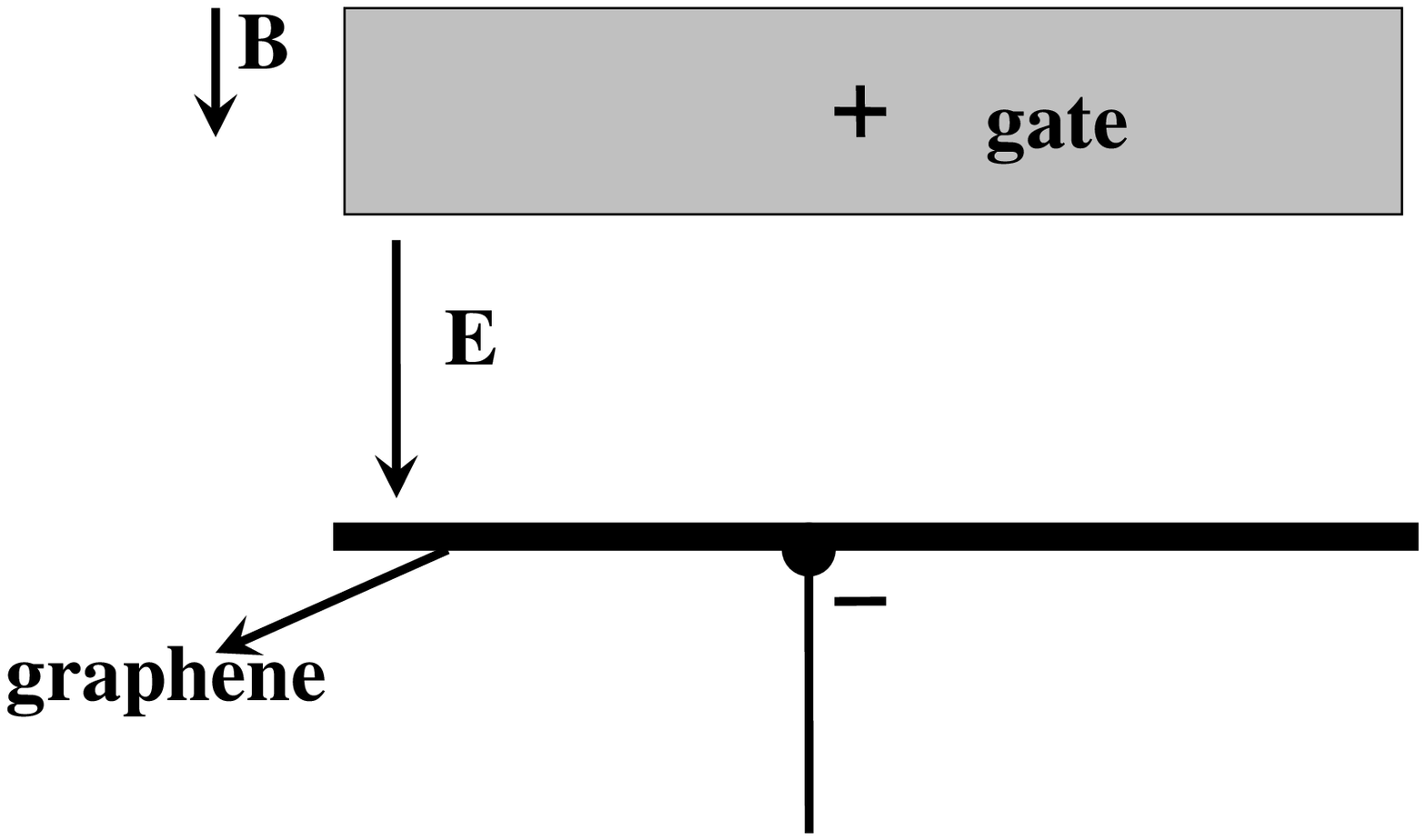}
   \caption{The graphene sheet in the presence of the applied electric $\mathbf{E}$ and magnetic $\mathbf{B}$ fields.}
   \label{fig_gr}
\end{figure}

When an undoped electron system in graphene in a magnetic field without an external electric field is in the ground state, half of the zeroth Landau level is filled with electrons, all Landau levels above the zeroth one are  empty, and all levels below  the zeroth one are filled with electrons. We suggest using the gate voltage shown in Fig.~\ref{fig_gr} to control the chemical potential in graphene by two ways: to shift it above the zeroth level so that it is between the zeroth and first Landau levels (the first case) or to shift the chemical potential below the zeroth level so that it is between the  first negative and zeroth Landau levels (the second case). In both cases, all Landau levels below the chemical potential are completely filled and all Landau levels above the chemical potential are completely empty. In the first case, there are allowed transitions between the zeroth and the first Landau levels, while in the second case there are allowed transitions between the first negative  and zeroth Landau levels (see the selection rules for optical transitions between the Landau levels in  single-layer graphene~[\onlinecite{Gusynin3}] and the analogous rules for the transitions between Landau levels in a 2D semiconductor~[\onlinecite{Ruvinsky2}]).
  Correspondingly, we consider magnetoexcitons formed in graphene by the electron on the first Landau level and the hole on the zeroth Landau level (the first case) or the electron on the zeroth Landau level and the hole on the Landau level $-1$ (the second case). Note that by appropriate gate potential we can also use  any other neighboring Landau levels $n$ and $n+1$.

It is obvious that magnetoexcitons formed in graphene are
two-dimensional, since graphene is a two-dimensional structure. Below we show that  for the relatively high dielectric constant of the microcavity,   $\epsilon \gg e^{2}/(\hbar v_{F}) \approx 2$ ($v_{F} =
\sqrt{3}at/(2\hbar)$  is the Fermi velocity of electrons in graphene, where
$a =2.566 \ \mathrm{\AA}$  is a lattice constant and $t \approx 2.71 \
\mathrm{eV}$ is the overlap integral between the nearest carbon
atoms \ \cite{Lukose}) the magnetoexciton energy in graphene can be
calculated by applying perturbation theory with respect to the strength of the
Coulomb electron-hole attraction analogously as it was done in [\onlinecite{Lerner}] for 2D quantum wells in a high magnetic field with non-zero electron
and hole masses ($m_{e} \neq 0$ and $m_{h} \neq 0$). This approach
allows us to obtain the spectrum
  of an isolated magnetoexciton with the electron on the Landau level $1$ and the hole on the Landau level $0$ in a single graphene layer.
The characteristic Coulomb electron-hole attraction for the single
graphene layer is $e^{2}/(\epsilon r_{B})$,  where $\epsilon $ is the dielectric constant of the
environment around graphene, $r_{B} = \sqrt{c\hbar/(eB)}$ denotes
the magnetic length of the magnetoexciton in the magnetic field $B$, and $c$ is the speed of light. The
energy difference between the first and zeroth Landau levels in
graphene is  $\hbar v_{F}/r_{B}$.    For graphene, the perturbative approach with respect to the strength of the Coulomb electron-hole attraction  is valid when $e^{2}/(\epsilon r_{B}) \ll \hbar
v_{F}/r_{B}$ \cite{Lerner}.
  This condition can be fulfilled at all magnetic
fields $B$ if the dielectric constant of the surrounding media satisfies the condition $
e^{2}/(\epsilon \hbar v_{F}) \ll 1$. Therefore, we claim that the
energy difference between the first and zeroth Landau levels is
always greater than the characteristic Coulomb attraction between the
electron and the hole in the single graphene layer at any $B$ if $\epsilon \gg e^{2}/(\hbar v_{F}) \approx 2$. Thus, applying perturbation theory with respect to weak Coulomb electron-hole attraction in graphene embedded in the $\mathrm{GaAs}$ microcavity ($\epsilon = 12.9$) is more accurate than for graphene embedded in the $\mathrm{SiO}_{2}$ microcavity ($\epsilon = 4.5$). This condition for perturbation theory in graphene is different from the 2D quantum well in GaAs, since in the latter case the energy deference between the neighboring Landau levels  is  $\hbar \omega_{c}$, where $\omega_{c} = e B/(c \mu_{eh}) $ is the cyclotron frequency, $\mu_{eh} = m_{e}m_{h}/(m_{e} + m_{h})$, and $m_{e}$ and $m_{h}$ are the effective masses of the electron and the hole, correspondingly \cite{Lerner}. Therefore, for the  quantum well in GaAs, the binding energy of the magnetoexciton is much smaller than the energy difference between two neighboring Landau levels only in the limit of high magnetic field $B \gg e^{3}c \mu_{eh}^{2}/(\epsilon^{2}\hbar^{3})$, and  perturbation theory with respect to weak electron-hole attraction can be applied only for high magnetic field.

 The operator for electron-hole Coulomb attraction is
\begin{eqnarray}\label{el-ho}
\hat{V}(r) = -\frac{e^{2}}{\epsilon r } \  ,
\end{eqnarray}
where $\mathbf{r} = \mathbf{r}_{e} - \mathbf{r}_{h}$, and
$\mathbf{r}_{e}$ and $\mathbf{r}_{h}$ are vectors of an electron and
a hole in a 2D plane, respectively.

A conserved quantity for an isolated electron-hole pair in a magnetic
field $B$ is the generalized magnetoexciton momentum
$\hat{\mathbf{P}}$ \cite{Gorkov,Lerner,Kallin}, which is given by
\begin{eqnarray}\label{Momentum}
\hat{\mathbf{P}} =  -i\hbar\nabla_{e} -i\hbar\nabla_{h} +
\frac{e}{c}(\mathbf{A}_{e} - \mathbf{A}_{h}) - \frac{e}{c}
[\mathbf{B}\times (\mathbf{r}_{e} -\mathbf{r}_{h})]\ .
\end{eqnarray}
The conservation of $\hat{\mathbf{P}}$ is related to the invariance of
the system upon the simultaneous translation of an electron and a hole
along with a gauge transformation. In Eq.~(\ref{Momentum}), the cylindrical
gauge for the vector potential is used: $\mathbf{A}_{e(h)} = 1/2
[\mathbf{B}\times \mathbf{r}_{e(h)}]$.

The eigenfunction  $\psi_{\tau}$ of the Hamiltonian  of the
two-dimensional electron-hole pair in graphene in the perpendicular
magnetic field $B$, which is also the eigenfunction of the generalized
momentum $\hat{\mathbf{P}}$,
 has the form~\cite{Gorkov,Lerner,Kallin}:

\begin{eqnarray}\label{psi}
\psi_{\mathbf{P}}(\mathbf{R},\mathbf{r}) = \exp\left[
i\left(\mathbf{P} + \frac{e}{2c}  [\mathbf{B}\times
\mathbf{r}]\right)
\frac{\mathbf{R}}{\hbar}\right]\tilde{\Phi}(\mathbf{r} -
\mathbf{\rho}_{0})\ ,
\end{eqnarray}
where $\mathbf{R}= (\mathbf{r}_{e} + \mathbf{r}_{h})/2$ and $\mathbf{\rho}_{0} =
c[\mathbf{B}\times\mathbf{P}]/(e B^{2})$.

The wave function of the relative coordinate
$\tilde{\Phi}(\mathbf{r})$ in Eq.~(\ref{psi}) can be expressed in
terms of the two-dimensional harmonic oscillator eigenfunctions
$\Phi_{n_{1},n_{2}}(\mathbf{r})$. For an electron at the Landau level
$n_{+}$ and a hole at the level $n_{-}$, the four-component wave
functions are \cite{Iyengar}

\begin{eqnarray}\label{electron}
\tilde{\Phi}_{n_{+},n_{-}}(\mathbf{r}) =   \left(
\sqrt{2}\right)^{\delta_{n_{+},0}+\delta_{n_{-},0}-2}\left(
\begin{array}{c} s_{+}s_{-}
\Phi_{|n_{+}|-1,|n_{-}|-1}(\mathbf{r})\\
s_{+}\Phi_{|n_{+}|-1,|n_{-}|}(\mathbf{r})\\
s_{-}\Phi_{|n_{+}|,|n_{-}|-1}(\mathbf{r})\\
\Phi_{|n_{+}|,|n_{-}|}(\mathbf{r})\ ,
\end{array}\right) ,
\end{eqnarray}
where $s_{\pm} = \mathrm{sgn} (n_{\pm})$.

In a high magnetic field, the magnetoexciton is formed by an electron on the Landau level $1$
and a hole on the  Landau level $0$ with the
 following four-component wave function:
\begin{eqnarray}\label{electron2}
\tilde{\Phi}_{1,0}(\mathbf{r}) =   \frac{1}{\sqrt{2}} \left(
\begin{array}{c}
0\\
\Phi_{0,0}(\mathbf{r})\\
0\\
\Phi_{1,0}(\mathbf{r})
\end{array}\right)\ ,
\end{eqnarray}
where $\Phi_{n_{1},n_{2}}(\mathbf{r})$ is the two-dimensional
harmonic oscillator eigenfunction given by
\begin{eqnarray}\label{electron_S}
\Phi_{n_{1},n_{2}}(\mathbf{r}) =
(2\pi)^{-1/2}2^{-|m|/2}\frac{\tilde{n}!}{\sqrt{n_{1}!n_{2}!}}
\frac{1}{r_{B}} \mathrm{sgn}(m)^{m}\frac{r^{|m|}}{r_{B}^{|m|}}
\exp\left[-im\phi - \frac{r^{2}}{4r_{B}^{2}}\right]
L_{\tilde{n}}^{|m|}\left(\frac{r^{2}}{2r_{B}^{2}}\right) \ .
\end{eqnarray}
In Eq.~(\ref{electron_S}), $L_{\tilde{n}}^{|m|}$  denotes Laguerre
polynomials, $m = n_{1} -n_{2}$, $\tilde{n} = \min(n_{1},n_{2})$,
and $\mathrm{sgn}(m)^{m} = 1$ for $m=0$. Note that we consider a magnetoexciton formed by an electron and a hole located  in the same type of  valley, e.g., in the point K (or K') of Brillouin zone.

 The magnetoexciton energies $E_{n_{+},n_{-}}(P)$ in graphene are functions of the generalized magnetoexciton momentum $\mathbf{P}$,  and in the first-order perturbation, are equal to
\begin{eqnarray}\label{en1}
E_{n_{+},n_{-}}(P) = E_{n_{+},n_{-}}^{(0)} +
\mathcal{E}_{n_{+},n_{-}}(P)\ .
\end{eqnarray}
In Eq.~(\ref{en1}), $E_{n_{+},n_{-}}^{(0)}$ is the energy of the electron-hole pair
when the electron is at the Landau level $n_{+}$ and the hole is at
the Landau level $n_{-}$, and it is given by \cite{Iyengar}
\begin{eqnarray}\label{energy2}
E_{n_{+},n_{-}}^{(0)} =  \frac{\hbar v_{F}}{r_{B}}
\sqrt{2}\left[\mathrm{sgn}(n_{+})\sqrt{|n_{+}|} -
\mathrm{sgn}(n_{-})\sqrt{|n_{-}|}\right]\ ,
\end{eqnarray}
while
\begin{eqnarray}\label{en2}
\mathcal{E}_{n_{+},n_{-}}(P) = - \left\langle
n_{+}n_{-}\mathbf{P}\left| \frac{e^{2}}{\epsilon r}
\right|n_{+}n_{-}\mathbf{P}\right\rangle \ ,
\end{eqnarray}
where $\left|n_{+}n_{-}\mathbf{P}\right\rangle \ =
\psi_{\mathbf{P}}(\mathbf{R},\mathbf{r})$ is defined by
Eq.~(\ref{psi}).

We calculate the magnetoexciton energy using the expectation value
of the electron-hole Coulomb attraction  for an electron on the Landau level $1$
and a hole on the  Landau level $0$. Neglecting the transitions between different Landau levels, the
first order perturbation with respect to the weak Coulomb attraction
results in the following expression for the energy of the
magnetoexciton:
\begin{eqnarray}\label{en3}
E_{1,0} (P) =  - \left\langle 1 \ 0 \
\mathbf{P}\left| \frac{e^{2}}{\epsilon|\mathbf{r}_{e} -
\mathbf{r}_{h}|} \right|1 \ 0 \ \mathbf{P}\right\rangle \ .
\end{eqnarray}
Denoting the averaging by the 2D harmonic oscillator eigenfunctions
$\Phi_{n_{1},n_{2}}(\mathbf{r})$ as $\langle\tilde{n}m
\mathbf{P}|\ldots| \tilde{n}m \mathbf{P}\rangle_{\Phi}$, where
$\tilde{n}$ and $m$ are defined below Eq.~(\ref{electron_S}), we get
the energy of a magnetoexciton created by the electron and hole on
the lowest Landau level:
\begin{eqnarray}\label{en4}
E_{1,0} (P) &=& \left\langle 1 \ 0\
\mathbf{P}\left|\hat{V}(r)\right|1 \ 0 \ \mathbf{P}\right\rangle  = \frac{1}{2}\left(
\left\langle 0 \ 0 \ \mathbf{P}\left|\hat{V}(r)\right|0 \ 0 \
\mathbf{P} \right\rangle_{\Phi}  + \left\langle 0 \ 1 \
\mathbf{P}\left|\hat{V}(r)\right|0 \ 1 \ \mathbf{P}
\right\rangle_{\Phi} \right) .
\end{eqnarray}
Following~[\onlinecite{Ruvinsky}], it is easy to show that, for small magnetic momenta $P \ll
\hbar/r_{B}$ for the electron-hole Coulomb attraction (\ref{el-ho}), each matrix element in Eq.~(\ref{en4}) can be expressed in terms of the binding energy and the effective magnetic mass of the magnetoexciton formed by an electron and a hole in the quantum well with the 2D electrons and holes:
\begin{eqnarray}\label{ruv}
\langle \tilde{n}m \mathbf{P}|\hat{V}(r)| \tilde{n}m
\mathbf{P}\rangle_{\Phi} = - \mathcal{E}_{\tilde{n}m}^{(b)} +
\frac{P^{2}}{2M_{\tilde{n}m}(B)} \ .
\end{eqnarray}
$\mathcal{E}_{\tilde{n}m}^{(b)}$ and $M_{\tilde{n}m}(B)$ are
the binding energy and the effective magnetic mass of the magnetoexciton, respectively,
corresponding to the magnetoexciton in the state with quantum
numbers $\tilde{n}$ and $m$.

Substituting Eq.~(\ref{ruv}) into Eq.~(\ref{en4}), we get the
dispersion law of a magnetoexciton for small magnetic momenta
\begin{eqnarray}\label{en55}
E_{1,0} (P) &=& \frac{1}{2}\left(\mathcal{E}_{00}^{(b)}(B) + \mathcal{E}_{01}^{(b)}(B)\right) +
\frac{1}{2}\left(\frac{1}{M_{00}(B)} +
 \frac{1}{M_{01}(B)} \right) \frac{P^{2}}{2}  .
\end{eqnarray}
Eq.~(\ref{en55}) can be rewritten in the form:
\begin{eqnarray}\label{en5}
E_{1,0} (P) &=&  - \mathcal{E}_{B}^{(b)} + \frac{P^{2}}{2m_{B}} \ ,
\end{eqnarray}
where the binding energy $\mathcal{E}_{B}^{(b)}$ and the effective
magnetic mass $m_{B}$ of a magnetoexciton  in  graphene with the electron on the Landau level $1$ and the hole on the Landau level $0$ are
\begin{eqnarray}\label{m1}
\mathcal{E}_{B}^{(b)} &=&  - \frac{1}{2} \left( \mathcal{E}_{00}^{(b)}(B) +  \mathcal{E}_{01}^{(b)}(B) \right) \ , \nonumber \\
\frac{1}{m_{B}} &=& \frac{1}{2} \left(  \frac{1}{M_{00}(B)} + \frac{1}{M_{01}(B)}\right) \ .
\end{eqnarray}
The constants $\mathcal{E}_{00}^{(b)}(B)$,
$\mathcal{E}_{01}^{(b)}(B)$,
$M_{00}(B)$, and $M_{01}(B)$ depend on the magnetic field
$B$, and are given in Ref.~[\onlinecite{Ruvinsky}]:
\begin{eqnarray}\label{m2}
\mathcal{E}_{00}^{(b)}(B) &=&
-\mathcal{E}_{0}\ , \nonumber \\
 \mathcal{E}_{01}^{(b)}(B) &=& -\frac{1}{2} \mathcal{E}_{0} \ , \nonumber \\
   M_{00}(B) &=&  M_{0} \ , \nonumber \\
 M_{01}(B) &=&  -2 M_{0} \ ,
 \end{eqnarray}
where $\mathcal{E}_{0}$  is the magnetoexcitonic energy and  $M_{0}$
is  the effective magnetoexciton mass in a quantum well. These
quantities are defined as
\begin{eqnarray}\label{m0}
\mathcal{E}_{0} &=& \sqrt{\frac{\pi}{2}} \frac{e^{2}}{\epsilon r_{B}} \ , \nonumber \\
M_{0} &=&  \frac{2^{3/2}\epsilon \hbar^{2}}{\sqrt{\pi}e^{2}r_{B}} \
.
\end{eqnarray}

Substituting Eq.~(\ref{m2}) into Eq.~(\ref{m1}) gives the binding
energy $\mathcal{E}_{B}^{(b)}$ and the effective magnetic mass
$m_{B}$ of the magnetoexciton  in a single graphene layer in a high
magnetic field:
\begin{eqnarray}\label{emsa}
\mathcal{E}_{B}^{(b)} =   \frac{3}{4} \mathcal{E}_{0} =
 \frac{3}{4} \sqrt{\frac{\pi}{2}}  \frac{e^{2}}{\epsilon r_{B}} ,
\hspace{3cm} m_{B} = 4 M_{0} = \frac{2^{7/2}\epsilon
\hbar^{2}}{\sqrt{\pi}e^{2}r_{B}} \ .
\end{eqnarray}
We can see that the effective magnetic mass of a 2D direct
magnetoexciton is  $4$ times higher in graphene than in a quantum well, while the magnetoexcitonic energy  is $3/4$ times lower in  graphene than in a quantum well at the same $\epsilon$ and $\mathbf{B}$. It is interesting to mention that we obtained the effective magnetic mass of the magnetoexciton in Eq.~(\ref{emsa}) using the four-component wavefunctions of magnetoexcitons in graphene given by Eqs.~(\ref{electron}) and~(\ref{electron2}). This reflects the specific and different properties of magnetoexcitons and, therefore, magnetopolaritons  in graphene compared to the polaritons in a quantum well  without a magnetic field~[\onlinecite{Berman_L_S}].

At small magnetic momentum ($P \ll \hbar/r_B$) for measuring
energies relative to the binding energy of a magnetoexciton,  the
dispersion relation $\varepsilon _{k}(P)$  of a magnetoexciton is
quadratic:
\begin{equation}\label{Energy}
\varepsilon _{k}({\bf P}) = \frac{P^2}{2m_{B k}} \ ,
\end{equation}
where $m_{B k}$ is the effective magnetic mass that  depends  on $B$
and the magnetoexcitonic quantum numbers $k = \{n_{+},n_{-}\}$ for
an electron at Landau level $n_{+}$ and a hole at level $n_{-}$.

It is easy to see that the results for the binding energy and effective magnetic mass of the exciton with the electron on the Landau level $0$ and the hole on the Landau level $-1$ will be exactly the same as for the exciton with the electron on the Landau level $1$ and the hole on the Landau level $0$.

We have derived above the spectrum of the single magnetoexciton in graphene (\ref{en5}), which is described by the eigenfunction of Dirac equation that has the four-component spinor structure given by Eq.~(\ref{electron2}). Alternatively, the wave function of the magnetoexciton in a QW has the one-component structure, because this wave function is the eigenfunction of Schr\"{o}dinger equation. However,  Eq.~(\ref{en5}) is valid also for a QW, but the binding energy and effective magnetic mass of 2D magnetoexciton formed by the electron and hole in the QW on the zeroth Landau level are given by \cite{Lerner}
\begin{eqnarray}\label{emsaqw}
\mathcal{E}_{B}^{(b)} =   \mathcal{E}_{0} =
 \sqrt{\frac{\pi}{2}}  \frac{e^{2}}{\epsilon r_{B}} ,
\hspace{3cm} m_{B} =  M_{0} = \frac{2^{3/2}\epsilon
\hbar^{2}}{\sqrt{\pi}e^{2}r_{B}} \ .
\end{eqnarray}
Also for a QW the expression for the single magnetoexciton spectrum given by Eq.~(\ref{Energy}) is valid.

\section{The effective Hamiltonian of trapped microcavity polaritons in graphene and in a QW in a high magnetic field}
\label{Hamil_eff}

Polaritons are linear superpositions of excitons and photons. In
high magnetic fields, when magnetoexcitons may exist, the polaritons
become linear superpositions of magnetoexcitons and photons. Let
us define the superpositions of magnetoexcitons and photons as
magnetopolaritons.  It is obvious that magnetopolaritons in graphene
are two-dimensional, since graphene is a two-dimensional structure.
The effective Hamiltonian of magnetopolaritons in  graphene and a QW in the strong magnetic field is
given by
\begin{eqnarray}
\label{Ham_tot_pol} \hat{H}_{tot} = \hat{H}_{mex} + \hat{H}_{ph} +
\hat{H}_{mex-ph} \ ,
\end{eqnarray}
where  $\hat{H}_{ph}$ is a photonic Hamiltonian, and $\hat{H}_{exc-ph}$ is the
Hamiltonian of magnetoexciton-photon interaction, and $\hat{H}_{mex}$ is a effective magnetoexcitonic Hamiltonian.
Let us analyze each term of the Hamiltonian for magnetopolaritons
(\ref{Ham_tot_pol}). It was shown in Ref.~[\onlinecite{Berman_K_L,Berman_K_L_2}] that 2D magnetoexcitons in graphene and a QW in a high magnetic field can be described by the same effective Hamiltonian $\hat{H}_{mex}$. The effective Hamiltonian
of $2D$ non-interacting magnetoexcitons in the infinite homogeneous
system in a high magnetic field is given by \cite{Berman_K_L,Berman_K_L_2}
\begin{eqnarray}
\label{Ham_exc} \hat H_{mex} = \sum_{{\bf P}}^{}\varepsilon_{mex}(P)
\hat{b}_{{\bf P}}^{\dagger}\hat{b}_{{\bf P}}^{} \ ,
\end{eqnarray}
 where $\hat{b}_{{\bf P}}^{\dagger}$ and $\hat{b}_{{\bf P}}$ are magnetoexcitonic creation
and annihilation operators obeying the Bose commutation relations.
For Hamiltonian (\ref{Ham_exc}),  the energy dispersion of a single
magnetoexciton in a graphene layer is given by
\begin{eqnarray}
\label{sp_exc} \varepsilon_{mex}(P) = E_{band} -
\mathcal{E}_{B}^{(b)} + \varepsilon _{0}(P) \ .
\end{eqnarray}
$E_{band}= E_{1,0}^{(0)} = \sqrt{2}\hbar v_{F}/r_{B}$ is the band
gap  energy, which is the difference between the Landau levels $1$ and $0$ in graphene defined by Eq.~(\ref{energy2}). $\mathcal{E}_{B}^{(b)}$ is the binding energy of a 2D
magnetoexciton with the electron in the Landau level $1$ and the hole on
the Landau level $0$ in a single graphene layer, and $\varepsilon
_{0}(P) = P^2/(2m_{B})$, where $m_{B}$ is the effective magnetic
mass of a 2D magnetoexciton with the electron on the Landau level
$1$ and hole on the Landau level $0$ in a single graphene layer
given by Eq.~(\ref{Energy}).

It can be shown that the interaction between two direct 2D
magnetoexcitons in graphene with the electron on the Landau level $1$ and the hole on the Landau level $0$ can be
neglected in a strong magnetic field, in analogy to what is described in Ref.~[\onlinecite{Lerner}] for 2D magnetoexcitons in a quantum well. The dipole moment
of each exciton in a magnetic field is $\mathbf{d}_{1,2} =
e\mathbf{\rho}_{0} =r_{B}^{2}\left[\mathbf{B}\times
\mathbf{P}_{1,2}\right]/B$ \cite{Lerner}, where
 $\mathbf{P}_{1}$ and $\mathbf{P}_{2}$ are the magnetic momenta of each exciton and $P_{1}, P_{2}\ll 1/r_{B}$. The magnetoexcitons are located
 at a distance $R \gg r_{B}$ from each other. The corresponding contribution to the energy of
  their dipole-dipole interaction is
   $\sim \mathcal{E}_{B}^{(b)} \left(r_{B}/R\right)^{3}P_{1}P_{2}r_{B}^{2}/\epsilon \sim \left(r_{B}/R\right)^{3}P_{1}P_{2}/(\epsilon M_{0}) \ll  e^{2}r_{B}^{2}/(\epsilon R^{3})$.
   Inputting the radius of the magnetoexciton in graphene   $r_{0,1} \sim r_{B}$
   \cite{Berman_L_G}, we obtain that
    the van der Waals attraction of
    the exciton at zero momenta is proportional to $\sim \left(r_{0,1}/R\right)^{6} \sim \left(r_{B}/R\right)^{6}$.
    Therefore, in the limit of a strong magnetic field for a dilute system
     $r_{B}\ll R$, both the dipole-dipole interaction and the van der Waals attraction vanish,
     and the 2D magnetoexcitons in graphene form an ideal Bose gas analogously to the 2D magnetoexcitons in a quantum well given in Ref.~[\onlinecite{Lerner}].
    Thus, the Hamiltonian (\ref{Ham_tot_pol}) does not include
     the term corresponding to the interaction between two direct
     magnetoexcitons in a single graphene layer.  So in high magnetic field  there is the BEC of the ideal magnetoexcitonic gas in
     graphene.

\begin{figure}[t] 
   \centering
  \includegraphics[width=3.5in]{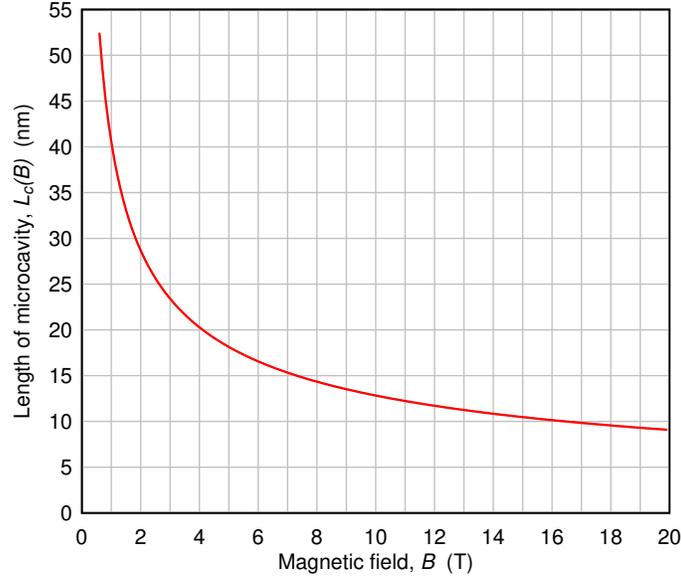}
   \caption{The length of the microcavity of GaAs ($\epsilon_{C} = 12.9$), corresponding to magnetoexciton-photon resonance,  as a function of the magnetic
   field $B$. }
   \label{fig_L}
\end{figure}

Let us analyze the other two terms in the Hamiltonian~(\ref{Ham_tot_pol}). The Hamiltonian of non-interacting photons in a semiconductor
microcavity is given by  \cite{Pau}:
\begin{eqnarray}
\label{Ham_ph} \hat H_{ph} = \sum_{{\bf P}}  \varepsilon _{ph}(P)
\hat{a}_{{\bf P}}^{\dagger}\hat{a}_{{\bf P}}^{} ,
\end{eqnarray}
 where $\hat{a}_{{\bf P}}^{\dagger}$ and $\hat{a}_{{\bf P}}$ are
photonic creation and annihilation Bose operators. The cavity
photon spectrum is given by
\begin{eqnarray}
\label{sp_phot} \varepsilon _{ph}(P) = (c/n)\sqrt{P^{2} +
\hbar^{2}\pi^{2}L_{C}^{-2}} \ .
\end{eqnarray}
In Eq.~(\ref{sp_phot}), $L_{C}$ is the length of the cavity, $n =
\sqrt{\epsilon_{C}}$ is the effective refractive index and
$\epsilon_{C}$ is the dielectric constant of the cavity. We assume that the
length of the microcavity has the following form:
\begin{eqnarray}
\label{lc} L_{C}(B) = \frac{\hbar\pi c}{n \left(E_{band} -
\mathcal{E}_{B}^{(b)}\right)} \ ,
\end{eqnarray}
corresponding to the resonance of the photonic and magnetoexcitonic branches at $P = 0$ (i.e.  $\varepsilon_{mex}(0) = \varepsilon_{ph}(0)$). The length of the microcavity, corresponding to a
magnetoexciton-photon resonance, decreases with the increment of
the magnetic field as $B^{-1/2}$. The dependence of the length of
the microcavity corresponding to the magnetoexciton-photon resonance on the
magnetic field is shown in Fig.~\ref{fig_L}. The resonance between
magnetoexcitons and cavity photonic modes can be achieved either by
controlling the spectrum  of magnetoexcitons $\varepsilon_{ex}(P)$
by changing magnetic field $B$ or by choosing  the appropriate length  of the
microcavity $L_{C}$. Let us mention that, while in the presence of a high magnetic field,  the length of the microcavity  corresponding to the magnetoexciton-photon resonance depends on the magnetic field as it is shown in Fig.~\ref{fig_L}. This effect does not take place in the system without a magnetic field~[\onlinecite{Berman_L_S}].

 The Hamiltonian of the harmonic
magnetoexciton-photon coupling has the form \cite{Ciuti}:
\begin{eqnarray}
\label{Ham_exph} \hat{H}_{mex-ph} = {\hbar \Omega_{R}}\sum_{{\bf P}}
 \hat{a}_{{\bf P}}^{\dagger}\hat{b}_{{\bf P}}^{} + h.c. ,
\end{eqnarray}
where the magnetoexciton-photon coupling energy represented by the
Rabi constant $\hbar \Omega_{R}$ is obtained in Sec.~\ref{Rabi}. Let us mention that $\Omega_{R}$ is obtained for a QW from the standard procedure describing  the electron-photon interaction in the Hamiltonian by the $\mathbf{P}\cdot \mathbf{A}$ term, while in a single graphene layer $\Omega_{R}$ is obtained from the electron-photon interaction based on the Dirac Hamiltonian for the electron in graphene.

 The excitonic and photonic operators are defined as~\cite{Ciuti}
\begin{eqnarray}
\label{bog_tr} \hat{b}_{\mathbf{P}} = X_{P}\hat{p}_{\mathbf{P}} -
C_{P}\hat{u}_{\mathbf{P}}, \hspace{3cm}
 \hat{a}_{\mathbf{P}} = C_{P}\hat{p}_{\mathbf{P}} +
X_{P}\hat{u}_{\mathbf{P}},
\end{eqnarray}
 where $\hat{p}_{\mathbf{P}}$ and $\hat{u}_{\mathbf{P}}$ are lower and upper magnetopolariton Bose
operators, respectively.  $X_{P}$ and $C_{P}$ are given by
\begin{eqnarray}
\label{bog} X_{P} = \frac{1}{\sqrt{1 +
\left(\frac{\hbar\Omega_{R}}{\varepsilon_{LP}(P) -
 \varepsilon _{ph}(P)}\right)^{2}}} , \hspace{3cm}
C_{P} = - \frac{1}{\sqrt{1 + \left(\frac{\varepsilon_{LP}(P) -
 \varepsilon _{ph}(P)}{\hbar\Omega_{R}}\right)^{2}}} \ ,
\end{eqnarray}
and the energy spectra of the lower/upper magnetopolaritons are
\begin{eqnarray}
\label{eps0} \varepsilon_{LP/UP}(P) &=& \frac{\varepsilon _{ph}(P)
+\varepsilon _{mex}(P)}{2}  \nonumber \\ &\mp&
\frac{1}{2}\sqrt{(\varepsilon_{ph}(P) - \varepsilon _{mex}(P))^{2} +
4|\hbar\Omega_{R}|^{2}} \ .
\end{eqnarray}
Eq.~(\ref{eps0}) implies a splitting of $2 \hbar \Omega_R$ between the upper and lower
states of polaritons at $P=0$, which is known as the
Rabi splitting. Let us also mention that $|X_{P}|^{2}$ and
$|C_{P}|^{2} = 1 - |X_{P}|^{2}$ represent the magnetoexciton and
cavity photon fractions in the lower magnetopolariton.

Substituting Eq.~(\ref{bog_tr}) into
Eqs.~(\ref{Ham_exc}),~(\ref{Ham_ph}) and~(\ref{Ham_exph}), we
conclude that the total Hamiltonian $\hat{H}_{tot}$
(\ref{Ham_tot_pol})  can be diagonalized by applying unitary
transformations~(\ref{bog_tr}) and has the form:
\begin{eqnarray}
\label{h0} \hat{H}_{tot} =
\sum_{\mathbf{P}}\varepsilon_{LP}(P)\hat{p}_{\mathbf{P}}^{\dagger}\hat{p}_{\mathbf{P}}
+\sum_{\mathbf{P}}\varepsilon_{UP}(P)\hat{u}_{\mathbf{P}}^{\dagger}\hat{u}_{\mathbf{P}},
\end{eqnarray}
where $\hat{p}_{\mathbf{P}}^{\dagger}$, $\hat{p}_{\mathbf{P}}$,
$\hat{u}_{\mathbf{P}}^{\dagger}$,  $\hat{u}_{\mathbf{P}}$ are the
Bose creation and annihilation operators for the lower and upper
magnetopolaritons, respectively.

Eq.~(\ref{h0}) is the Hamiltonian of magnetopolaritons in a single
graphene layer in a high magnetic field. Our particular interest is
the lower energy magnetopolaritons which produce the BEC. The lower
palaritons have the lowest energy within a single graphene layer.
Therefore, from Eq.~(\ref{h0}) we can obtain
\begin{eqnarray}
\label{Ham_tot_p} \hat H_{tot} =
\sum_{\mathbf{P}}\varepsilon_{LP}(P)\hat{p}_{\mathbf{P}}^{\dagger}\hat{p}_{\mathbf{P}}
 \ .
\end{eqnarray}

Similarly to the case of Bose atoms in a
trap \cite{Pitaevskii,Mullin} in the case of a slowly varying
external potential, we can make the quasiclassical approximation,
 assuming that the effective magnetoexciton mass does not depend on a characteristic size $l$  of the trap and it is a constant within the trap.
  This quasiclassical approximation is valid if $P \gg \hbar/l$. The harmonic trap is formed by the two-dimensional planar potential in the plane of graphene.
  The potential trap can be produced in two
different ways. In case 1, the potential trap can be produced by
applying an external inhomogeneous electric field or inhomogeneous local stress. The spatial
dependence of the external field potential $V(r)$ is caused by
shifting of  magnetoexciton energy by applying an external
inhomogeneous electric field or inhomogeneous local stress. The photonic states in the cavity are
assumed to be unaffected by this electric field or stress. In this case the
band energy $E_{band}$ is replaced by $E_{band}(r) = E_{band}(0)+
V(r)$. Near the minimum of the magnetoexciton energy, $V(r)$ can be
approximated by the planar harmonic potential $\gamma r^{2}/2 $,
where $\gamma$ is the spring constant. Note that a high magnetic field
does not change the
 trapping potential in the effective Hamiltonian \cite{Ruvinsky2,Berman_L_S_C}. In
case 2, the trapping of magnetopolaritons is caused by the
inhomogeneous shape of the cavity when the length of the cavity is
given by
\begin{eqnarray}
\label{lc11}
L_{C}(r) = \frac{\hbar\pi c}{n\left(E_{band} -
\mathcal{E}_{B}^{(b)} + \gamma r^{2}/2 \right)} \ ,
\end{eqnarray}
where $r$ is the distance between the photon
 and the center of the trap. In case 2, the $\gamma$ in Eq.~(\ref{lc11}) is the curvature characterizing the shape of the cavity.  In case 1, for the slowly changing confining potential $V(r) =
\gamma  r^{2}/2$, the magnetoexciton
spectrum is    given in the effective mass approximation as
\begin{eqnarray}
\label{ex_sp} \varepsilon_{mex}^{(0)}(P) = \varepsilon_{mex}(P) +
V(r) = (c/n) \hbar\pi L_{C}^{-1} +  \frac{\gamma}{2}r^{2} +
\frac{P^{2}}{2m_{B}} \ ,
\end{eqnarray}
where $r$ is now the distance between the center of mass of
the   magnetoexciton and the center of the trap.
The Hamiltonian for photons in this case is given by
Eq.~(\ref{Ham_ph}), the spectrum of photons is shown by
Eq.~(\ref{sp_phot}) and the length of the microcavity is given by
Eq.~(\ref{lc}).

In case 2, for the slowly changing shape of the length of cavity given by Eq.~(\ref{lc11}), the photonic spectrum is given in the effective mass approximation as
\begin{eqnarray}
\label{ph_sp} \varepsilon _{ph}^{(0)}(P) = (c/n)\sqrt{P^{2} +
\frac{n^{2}}{c^{2}}\left(E_{band} - \mathcal{E}_{B}^{(b)} +
\frac{\gamma r^{2}}{2} \right)} \ .
\end{eqnarray}
This quasiclassical approximation is valid if $P \gg \hbar /l$,
where  $l = \left(\hbar/(m_{B}\omega_{0})\right)^{1/2}$ is the size
of the magnetoexciton cloud in an ideal magnetoexciton gas and
$\omega_{0} = \sqrt{\gamma/m_{B}}$. The Hamiltonian and spectrum of
magnetoexcitons in this case are given by Eq.~(\ref{Ham_exc})
and~(\ref{sp_exc}), correspondingly.

The  total Hamiltonian $\hat{H}_{tot}$ can be diagonalized by
applying unitary transformations. At small momenta $\alpha \equiv
1/2 (m_{B}^{-1} + (c/n)L_{C}/\hbar\pi)P^{2}/|\hbar \Omega_{R}| \ll
1$ ( $L_{C} = \hbar\pi c/n \left(E_{band} -
\mathcal{E}_{B}^{(b)}\right)^{-1}$) and weak confinement $\beta
\equiv \gamma r^{2}/|\hbar \Omega_{R}| \ll 1$, the single-particle
lower magnetopolariton spectrum obtained through the substitution of
Eq.~(\ref{ex_sp}) into Eq.~(\ref{eps0}), in linear order with
respect to the small parameters $\alpha$ and $\beta$, is
\begin{eqnarray}
\label{eps00} \varepsilon_{0}(P) \approx \frac{c}{n} \hbar \pi
L_{C}^{-1} - |\hbar \Omega_{R}| +  \frac{\gamma}{4} r ^{2} +
\frac{1}{4} \left(m_{B}^{-1} + \frac{c
L_{C}(B)}{n\hbar\pi}\right)P^{2} .
\end{eqnarray}
Let us emphasize that the spectrum of non-interacting
magnetopolaritons $\varepsilon_{0}(P)$  at small momenta and weak
confinement is given by Eq.~(\ref{eps00}) for  both physical
realizations of confinement: case 1 and case 2. By substituting
Eq.~(\ref{ex_sp}) into Eq.~(\ref{bog}), we obtain $X_{\mathbf{P}}
\approx 1/\sqrt{2}$. The condition for the validity of the
quasiclassical approach in Eq.~(\ref{Ham_exc}), $Pl\gg\hbar$, is
also applied here.

\begin{figure}[t] 
   \centering
  \includegraphics[width=3.5in]{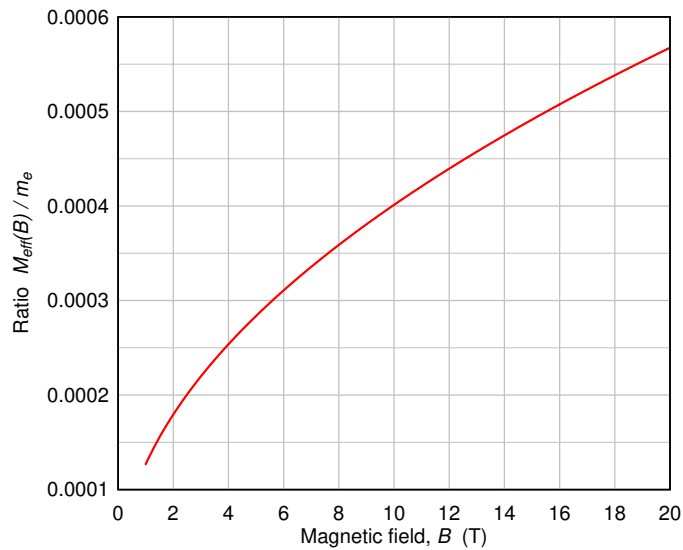}
   \caption{The ratio of the effective magnetopolariton mass $M_{\rm eff}(B)$ to the mass of a free electron $m_{e}$
  as a function of magnetic field $B$. }
   \label{fig_M}
\end{figure}

If we measure the energy relative to the $P=0$ lower magnetopolariton
energy $(c/n) \hbar \pi L_{C}^{-1}  - |\hbar \Omega_{R}|$, we obtain
the resulting  effective Hamiltonian for trapped magnetopolaritons
in graphene in a magnetic field. At small momenta $\alpha \ll 1$
($L_{C} = \hbar\pi c/n \left(E_{band} -
\mathcal{E}_{B}^{(b)}\right)^{-1}$) and weak confinement $\beta
 \ll 1$, this effective Hamiltonian is
\begin{eqnarray}
\label{Ham_eff} \hat H_{\rm eff}  =
\sum_{\mathbf{P}}\left(\frac{P^{2}}{2M_{\rm eff}(B)} + \frac{1}{2}
V(r) \right)\hat{p}_{\mathbf{P}}^{\dagger}\hat{p}_{\mathbf{P}} \ ,
\end{eqnarray}
 where the sum over $\mathbf{P}$  is carried out only over $P \gg \hbar/l$
  (only in this case the quasiclassical approach  used in Eq.~(\ref{ex_sp}) is valid), and
 the effective magnetic mass of a magnetopolariton is
given by
\begin{eqnarray}
\label{Meff} M_{\rm eff}(B) = 2   \left(m_{B}^{-1} + \frac{c
L_{C}(B)}{n\hbar\pi}\right)^{-1} \ .
\end{eqnarray}
 According to Eq.~(\ref{Meff}), the effective
magnetopolariton mass $M_{\rm eff}$ increases with the increment of
the magnetic field as $B^{1/2}$, as shown in Fig.~\ref{fig_M}.
 Let us emphasize that the resulting effective Hamiltonian for magnetopolaritons  in graphene in a magnetic field for the parabolic trap is given by
Eq.~(\ref{Ham_eff}) for both physical realizations of
confinement represented by case 1 and case 2. The effective magnetic mass of the magnetopolariton in a QW is approximately the same as in graphene, since the contribution to $M_{\rm eff}(B)$ from the second term in the r.h.s. of Eq.~(\ref{Meff})  is much higher than from the first term. So  the effective mass of the magnetopolariton in a QW can also be presented by Fig.~\ref{fig_M}.

Let us mention that the  effective Hamiltonian of magnetopolaritons in a QW in microcavity is also given by Eq.~(\ref{Ham_eff}) with the effective magnetic mass of magnetoexciton with the  electron and hole on the zeroth Landau level provided by Eq.~(\ref{emsaqw}).


\section{The Rabi splitting constant in graphene and a QW in high magnetic
field}

\label{Rabi}

Neglecting anharmonic terms for the magnetoexciton-photon coupling, the Rabi splitting constant $\Omega_{R}$ can be estimated
quasiclassically as
\begin{eqnarray}
\label{defrabi}
\left\vert \hbar\Omega_{R}\right\vert =\left\vert \left\langle f\left\vert
\hat{H}_{int}\right\vert i\right\rangle \right\vert  \ ,
\end{eqnarray}
where $\hat{H}_{int}$ is the Hamiltonian of the electron-photon
interaction. For graphene this interaction is determined by Dirac electron Hamiltonian as
\begin{eqnarray}
\label{graphint}
\hat{H}_{int}=-\frac{v_{F}e}{c}\vec{\hat{\sigma}}\cdot\vec{A}_{ph0} =\frac{v_{F}%
e}{i\omega}\vec{\hat{\sigma}}\cdot\vec{E}_{ph0} \ ,
\end{eqnarray}
where $\vec{\hat{\sigma}}=(\hat{\sigma}_{x},\hat{\sigma}_{y})$, $\hat{\sigma
}_{x}$ and $\hat{\sigma}_{y}$ are Pauli matrices, $\vec{A}_{ph0}$  is the vector potential corresponding to a single cavity photon, and $\ E_{ph0}=\left(8\pi\hbar\omega/(\epsilon W)\right)^{1/2} \ $ is the magnitude of
electric field corresponding to a single cavity photon of the frequency
$\omega$ in the volume of microcavity $W$, while for the QW this interaction
is %
\begin{eqnarray}
\label{hamintqw}
\hat{H}_{int}=\overrightarrow{d}_{12}\cdot\vec{E}_{ph0} \ ,
\end{eqnarray}
where
\begin{eqnarray}
\label{dip}
\overrightarrow{d}_{12} = e \sum_{i}\mathbf{r}_{i}
\end{eqnarray}
is the dipole momentum of transition and the sum is taken over the coordinate vectors related to the positions of all the electrons in the system.

In Eq.~(\ref{defrabi}) the initial $|i\rangle$ and final $|f\rangle$ electron states are different for graphene and a quantum well. For the case of graphene these electron states are
defined as
\begin{eqnarray}
\label{states}
| i \rangle &=& \prod_{k}\hat{c}_{0,k}^{\dagger} | 0 \rangle_{0} | 0 \rangle_{1}  \  ,  \nonumber \\
| f \rangle &=& \hat{b}_{1,0}^{\dagger} | i \rangle   \  .
\end{eqnarray}
In Eq.~(\ref{states}), $\hat{c}_{n,k}^{\dagger}$ is the Fermi creation operator of the electron with the $y$ component of the wavevector $k$ on the Landau level $n$, $| 0 \rangle_{n}$ denotes the wavefunction of the vacuum on the Landau level $n$, $\prod_{k}\hat{c}_{0,k}^{\dagger} | 0 \rangle_{0}$ corresponds to the completely filled zeroth Landau level, $\hat{b}_{n,n^{\prime}}^{\dagger}$ is the Bose creation operator of the magnetoexciton with the electron on the Landau level $n$ and the hole on the Landau level $n^{\prime}$. We consider magnetoexcitons with magnetic momenta equal to zero, for which  the Bose condensate in the system of non-interacting particles is the exact solution of the problem~\cite{Lerner}. Following Ref.~\cite{Lerner} $\hat{b}_{n,n^{\prime}}^{\dagger}$  for this case is defined as
\begin{eqnarray}
\label{opex}
\hat{b}_{n,n^{\prime}}^{\dagger} = \frac{1}{\sqrt{N_{d}}} \sum_{k} \hat{h}_{n^{\prime},k}^{\dagger} \hat{c}_{n,-k}^{\dagger}  \  ,
\end{eqnarray}
where  $\hat{h}_{n^{\prime},k}^{\dagger}$ is the Fermi creation operator of the hole  with the $y$ component of the wavevector $k$ on the Landau level $n^{\prime}$, $N_{d} = S/(2\pi r_{B}^{2})$ is  the macroscopic degeneracy of Landau levels, and $S$ is the area of the system.

Let us use the Landau gauge for the wavefunction of the single electron $\psi_{n,k}(x,y)$ with the $y$ component of the wavevector $k$ on the Landau level $n$.
In the Landau gauge with the vector potential $\mathbf{A} = (0,Bx,0)$, the two-component eigenfunction  $\psi_{n,k}(\mathbf{r})$
is given  by \cite{Ando}
\begin{eqnarray}
\label{efunction}
\psi_{n,k}(x,y) = \frac{C_{n}}{\sqrt{L_{y}}}\exp(ik y)
\left( \begin{array}{c} s(n) i^{n-1}\Phi_{n-1}(x-r_B ^{2}k ) \\
i^{n}\Phi_{n}(x-r_B ^{2}k)
\end{array}\right)\ ,
\end{eqnarray}
where  $s(n)$ defined by
\begin{eqnarray}
\label{sn}
s(n) = \left\{ \begin{array}{cc} 0& (n=0)\ ,  \\
 \pm 1 & (n>0)\ .
\end{array}\right.
\end{eqnarray}
 $L_y$ are normalization lengths in the  $y$ direction,
\begin{eqnarray}
\label{Cn}
C_n = \left\{ \begin{array}{cc} 1& (n=0)\ ,  \\
1/\sqrt{2} & (n>0)\ ,
\end{array}\right.
\end{eqnarray}
and
\begin{eqnarray}
\label{phin}
\Phi_n(x) = \left(2^{n}n!\sqrt{\pi}r_B \right)^{-1/2}
\exp\left[-\frac{1}{2}\left(\frac{x}{r_B }\right)^{2}\right]
H_{n}\left(\frac{x}{r_B }\right)\ ,
\end{eqnarray}
where $H_n(x)$ is the  Hermite polynomial. The corresponding eigenenergies depend
 on the quantum number $n$ only and are given by
\begin{eqnarray}
\label{energy0}
\varepsilon_{n} = \frac{\hbar v_F }{r_B } \sqrt{2n}\ .
\end{eqnarray}

Substituting Eqs.~(\ref{opex}) and~(\ref{efunction})  into~(\ref{states}) and using the electron-photon interaction $\hat{H}_{int}$~(\ref{graphint}), we finally obtain from Eq.~(\ref{defrabi}):
\begin{eqnarray}
\label{dipxy}
\left|\hbar \Omega_{R}\right|&=& \left|\frac{e v_{F}}{i \omega} \int dx \int dy x \left[\psi_{1,k}^{*}(x,y)  \vec{\hat{\sigma}}\cdot \vec{E}_{ph0} \psi_{0,k}(x,y)\right] \right| = \frac{ev_{F} \left|E_{ph0} \right|}{\sqrt{2}\omega} \ .
\end{eqnarray}

In Eq.~(\ref{dipxy}) the energy of photon  absorbed at the creation of the magnetoexciton (at $\mathcal{E}_{B}^{(b)} \ll \varepsilon_{1} - \varepsilon_{0}$) is given by
\begin{eqnarray}
\label{energy00}
\hbar \omega = \varepsilon_{1} - \varepsilon_{0}  = \sqrt{2} \frac{\hbar v_F }{r_B } \ .
\end{eqnarray}
Substituting the photon energy from Eq.~(\ref{energy00}) into   Eq.~(\ref{dipxy}), we obtain the Rabi splitting corresponding to the creation of a magnetoexciton with the electron on the Landau level $1$ and the hole on the Landau level $0$ in graphene:
\begin{eqnarray}
\label{rabigraph}
\hbar \Omega_{R} = 2 e \left(\frac{\pi \hbar v_F r_{B}}{\sqrt{2} \epsilon W} \right)^{1/2}  \ .
\end{eqnarray}
As follows from Eq.~(\ref{rabigraph}), the Rabi splitting in graphene is related to the creation of the magnetoexciton, which decreases when the magnetic field increases and is proportional to $B^{-1/4}$. Therefore, the Rabi splitting in graphene can be controlled by the external magnetic field. Note that in a semiconductor quantum well contrary to graphene the  Rabi splitting does not depend on the magnetic field.

Substituting Eq.~(\ref{hamintqw}) and  the initial $|i\rangle$ and final $|f\rangle$ electron states from Ref.~[\onlinecite{Lerner}] into~(\ref{defrabi})  after the integration we obtain  the Rabi splitting constant $\Omega_{R}$  for a quantum well
\begin{eqnarray}
\label{defrabiqw}
\hbar \Omega_{R} = d_{12}E_{ph0} \ ,
\end{eqnarray}
where $d_{12}$ is the matrix term of a magnetoexciton generation
transition in a QW  represented as
\begin{eqnarray}
\label{dipo}
d_{12} = e\left|\left\langle  f \left|\sum_{i}\mathbf{r}_{i}  \right| i \right \rangle  \right| \ .
\end{eqnarray}

The similar calculations for the transition dipole moment and the photon energy corresponding to the formation of magnetoexciton with the electron and hole on zeroth Landau level in the QW gives:
\begin{eqnarray}
\label{energy000}
d_{12} &=& \frac{e r_{B}}{2\sqrt{2}} \ , \nonumber \\
\hbar \omega &=& \varepsilon_{1} - \varepsilon_{0}  = \hbar \omega_{c} = \frac{\hbar e B}{c \mu_{eh}} \ .
\end{eqnarray}
 Substituting the transition dipole moment and the photon energy given by Eq.~(\ref{energy000}) into Eq.~(\ref{defrabi}), we obtain the Rabi splitting for QW:
\begin{eqnarray}
\label{rabiqw}
\hbar \Omega_{R} = 2 e \hbar \left(\frac{\pi}{\epsilon \mu_{eh} W} \right)^{1/2}   \ .
\end{eqnarray}
Thus, as it follows from Eq.~(\ref{rabiqw}), the Rabi splitting in a QW does not depend on the magnetic field in the limit of high magnetic field. Therefore, only in graphene can the Rabi splitting  be controlled by the external magnetic field in the limit of high magnetic field.

It is easy to show that the Rabi  splitting related to the creation of the magnetoexciton, the electron on the Landau level $0$ and the hole on the Landau level $-1$ will be exactly the same as for the magnetoexciton with the electron on the Landau level $1$ and the hole on the Landau level $0$.
Let us mention that dipole optical transitions from the Landau level $-1$ to the  Landau level $0$, as well as from  the Landau level $0$ to the  Landau level $1$, are allowed by the selection rules for optical transitions in single-layer
graphene \cite{Gusynin3}.


\section{Bose-Einstein condensation of  trapped microcavity magnetopolaritons in graphene and QW}
\label{bec}

Although Bose-Einstein condensation cannot  take place in a 2D
homogeneous ideal gas at non-zero temperature, as discussed in
Ref.~[\onlinecite{Bagnato}], in a harmonic trap the BEC can occur in two
dimensions below a critical temperature $T_{c}^{0}$. Below we estimate this temperature. In a harmonic
trap at a temperature $T$ below a critical temperature $T_{c}^{0}$ ($T
< T_{c}^{0}$), the number $N_{0}(T,B)$ of non-interacting magnetopolaritons in
the condensate  is given by \cite{Bagnato}
\begin{eqnarray}
 \label{n_con}
N_{0}(T,B) &=&  N - \frac{\Gamma(2)\zeta(2)
 \left(g_{s}^{(e)}g_{v}^{(e)} +
 g_{s}^{(h)}g_{v}^{(h)}\right)M_{\rm eff}(B)}{\hbar^{2}\gamma_{\rm eff}}(k_{B}T)^{2}
 \nonumber \\
&=& N - \frac{\pi
 \left(g_{s}^{(e)}g_{v}^{(e)} +
 g_{s}^{(h)}g_{v}^{(h)}\right)M_{\rm eff}(B)}{3\hbar^{2}\gamma}(k_{B}T)^{2} \ ,
\end{eqnarray}
where $N$ is the total number of magnetopolaritons,
$g_{s}^{(e),(h)}$ and $g_{v}^{(e),(h)}$ are the spin and graphene
valley degeneracies for an electron and a hole, respectively,
$k_{B}$ is the Boltzmann constant, $\Gamma(x)$ is the gamma function
and $\zeta (x)$ is the Riemann zeta function.

Applying the condition $N_{0}=0$ to Eq.~(\ref{n_con}), and assuming
that the magnetopolariton effective mass  is given by Eq.~(\ref{Meff}),
we obtain the BEC critical temperature $T_{c}^{(0)}$ for the ideal
gas of magnetopolaritons in a single graphene layer  in a magnetic
field:
\begin{eqnarray}
 \label{t_c}
T_{c}^{(0)} (B) = \frac{1}{k_{B}}\left(\frac{3\hbar^{2}\gamma N}{\pi
 \left(g_{s}^{(e)}g_{v}^{(e)} +
 g_{s}^{(h)}g_{v}^{(h)}\right)M_{\rm eff}(B)}
\right)^{1/2} \ .
\end{eqnarray}
At temperatures above $T_{c}^{(0)}$,  the BEC of magnetopolaritons in a single graphene layer does not exist.

\begin{figure}[t] 
   \centering
  \includegraphics[width=3.5in]{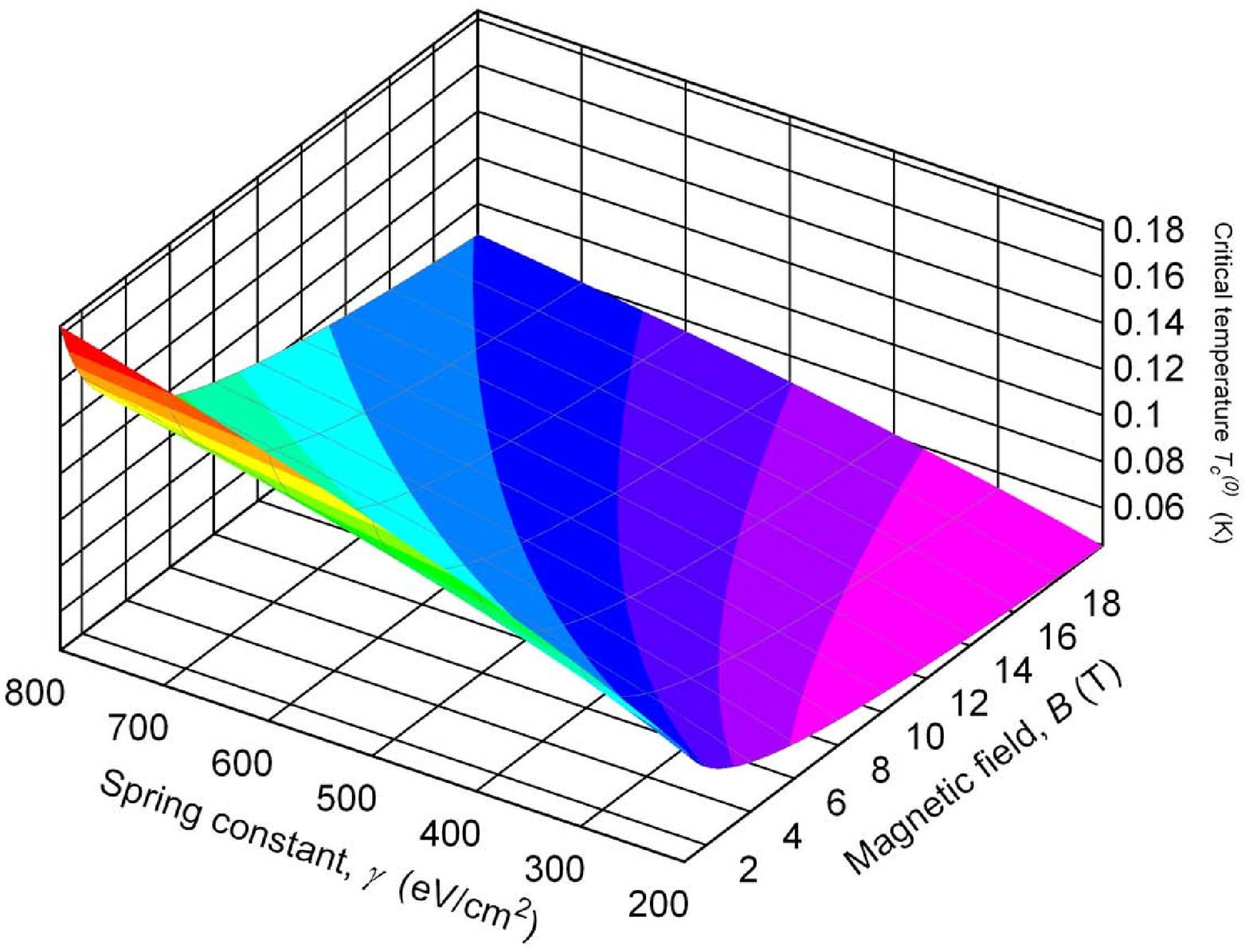}
\caption{The ratio of the BEC critical temperature to the square
root of the total number of magnetopolaritons $T_{c}^{(0)}/\sqrt{N}$
as a function of the magnetic
   field $B$  and the pring constant $\gamma$. We assume that the environment around graphene is $\mathrm{GaAs}$ with $\epsilon = 12.9$.}
   \label{3DT}
\end{figure}

\begin{figure}[t] 
   \centering
  \includegraphics[width=3.5in]{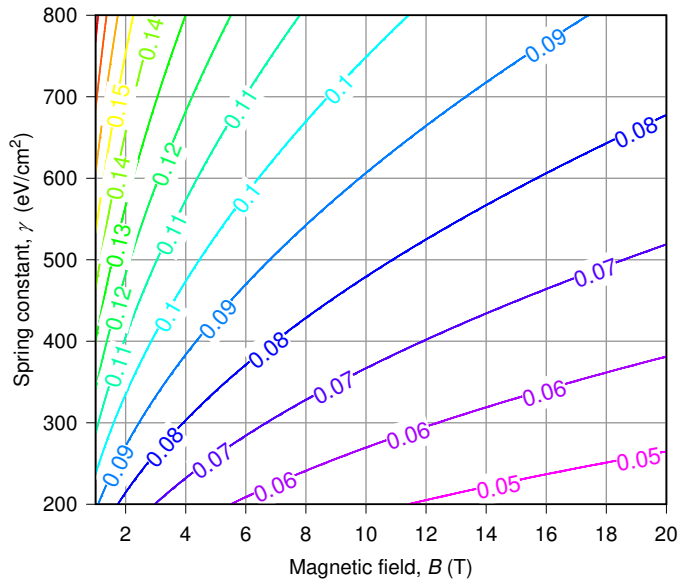}
   \caption{The functional relations between the spring constant $\gamma$ and
the magnetic    field $B$  corresponding to the different constant
values of $T_{c}^{(0)}/\sqrt{N}$. We assume that the environment around
graphene is $\mathrm{GaAs}$ with $\epsilon = 12.9$.}
   \label{2DT}
\end{figure}

\begin{figure}[t] 
   \centering
  \includegraphics[width=3.5in]{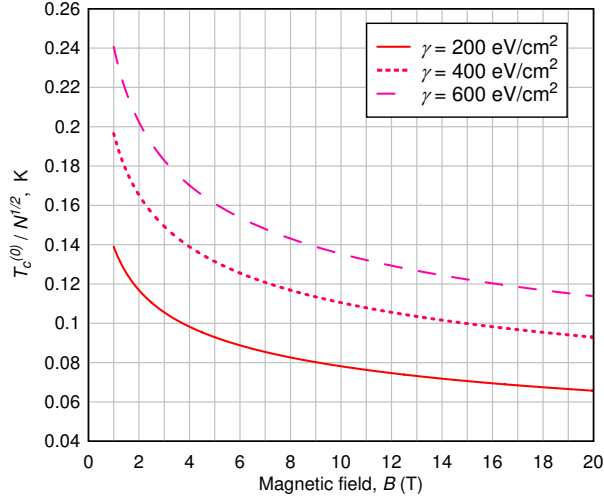}
   \caption{The ratio of the BEC critical temperature to the square root of the total number of magnetopolaritons $T_{c}^{(0)}/\sqrt{N}$ as a function of magnetic
   field $B$  at different spring constants $\gamma$. We assume the environment around graphene is $\mathrm{GaAs}$ with $\epsilon = 12.9$.}
   \label{fig_t_c}
\end{figure}

A three-dimensional plot of
$T_{c}^{(0)}/\sqrt{N}$ as a function of magnetic
   field $B$  and spring constant $\gamma$ is presented in
   Fig.~\ref{3DT}. In our calculations, we used $g_{s}^{(e)} = g_{v}^{(e)} =
g_{s}^{(h)} = g_{v}^{(h)} = 2$. The functional relations between the
spring constant $\gamma$ and the magnetic    field $B$ corresponding to
 different constant values of $T_{c}^{(0)}/\sqrt{N}$ are
presented in Fig.~\ref{2DT}.
 According to Eq.~(\ref{t_c}), the BEC critical temperature
$T_{c}^{(0)}$ decreases with the magnetic field  as
$B^{-1/4}$ and increases with  the spring constant
as $\gamma^{1/2}$. These functional relations are
illustrated  in Figs.~\ref{3DT}, \ref{2DT} and~\ref{fig_t_c}.

Substituting Eq.~(\ref{t_c}) into Eq.~(\ref{n_con}), we obtain
\begin{eqnarray}
 \label{t_c1}
\frac{N_{0}(T,B)}{N} = 1 - \left(\frac{T}{T_{c}^{(0)}(B)}
\right)^{2} \ .
\end{eqnarray}

Note that, since the quadratic spectrum of non-interacting
magnetopolaritons given by Eq.~(\ref{Energy}) does not satisfy
the Landau criterion of superfluidity \cite{Abrikosov,Griffin}, the
ideal Bose gas of magnetopolaritons in high magnetic field in
graphene is not a superfluid.

Since magnetopolaritons in a QW are described by the same effective Hamiltonian as in graphene, but with the different magnetic mass of the magnetoexciton, the results of the calculations presented in Figs.~\ref{3DT},~\ref{2DT} and~\ref{fig_t_c} for the critical temperature of the BEC for magnetopolaritons in graphene are valid for the BEC in a QW in a high magnetic field. This is true because the contribution to  the effective mass of the magnetopolariton from the second term in the r.h.s. of Eq.~(\ref{Meff}) is much higher than from the first term.


\section{Discussion and conclusions}
\label{discussion}

In our calculations, we have assumed that the system under consideration is in thermal equilibrium. This assumption  is valid if   the relaxation time is less than the quasiparticle lifetime. Although  the
magnetopolariton lifetime is short,
 thermal equilibrium can be achieved within the regime of a
strong pump.   Porras et al. \cite{Porras}
claimed that the time scale for polariton-exciton scattering can
be small enough to satisfy this condition for the existence of  a thermalized distribution of polaritons
in the lowest $k$-states in a quantum well. We expect a similar
characteristic time for magnetopolariton-magnetoexciton scattering
in graphene.  However, the consideration of pump and decay in a steady state may lead to results which are different from the ones presented in this paper. The consideration of the influence of decay on the BEC may be the
subject of further studies of a trapped gas.

Above we discussed the BEC of the magnetopolaritons in a single
graphene layer placed within a strong magnetic field. What would happen in  a
multilayer graphene system in a high magnetic field? Let us mention
that the magnetopolaritons formed by the microcavity photons and the
indirect excitons with the spatially separated electrons and holes
in different parallel graphene layers embedded in a semiconductor
microcavity can exist only at very low temperatures $k_{B} T \ll
\hbar \Omega_{R}$. For the case of the spatially separated electrons
and holes, the Rabi splitting $\Omega_{R}$ is very small in comparison to
the case of electrons and holes placed in a single graphene layer. This is
because $\Omega_{R} \sim d_{12}$ and the  matrix element of
magnetoexciton generation transition $d_{12}$ is proportional to the
overlapping integral of the electron and hole wavefunctions, which
is very small if the electrons and holes are placed in different
graphene layers. Therefore, we cannot predict the effect of
relatively high BEC critical temperature for the electrons and holes
placed in different graphene layers.

Spin polarization is important not only for the excitations but
for the condensate itself. It was shown in~[\onlinecite{Kavokin2,Kavokin3}] that taking into account the
spin degree of freedom can qualitatively modify the results for
exciton-magnetopolariton condensation at magnetic fields lower than
the critical magnetic field. We assume that magnetic field $B$
under consideration is above the critical one and, therefore,
the Zeeman splitting does not affect the spectrum of collective
excitations according to Fig.~1 in~[\onlinecite{Kavokin2}].
 So we neglect the Zeeman splitting in our calculations.

To conclude, we have derived  the effective Hamiltonian of the ideal
gas of trapped cavity magnetopolaritons in a single graphene layer and a QW
in a high magnetic field. The resonance between magnetoexcitons and
cavity photonic  modes can be achieved either by  controlling the
spectrum  of magnetoexcitons $\varepsilon_{ex}(P)$, by changing
magnetic field $B$ or by controlling the length  of  the microcavity
$L_{C}$.  We analyzed two possible physical realizations of the
trapping potential: inhomogeneous local stress or a harmonic electric field potential coupled to
magnetoexcitons and a parabolic shape of the semiconductor cavity
causing the trapping of microcavity photons. We conclude that both
 realizations of  confinement result in the same effective
Hamiltonian. It is shown that the effective magnetopolariton mass
$M_{\rm eff}$ increases with the  magnetic field as
$B^{1/2}$. Meanwhile, the BEC critical temperature $T_{c}^{(0)}$
decreases as $B^{-1/4}$ and increases with the spring constant
 as $\gamma^{1/2}$.  The gas of magnetopolaritons in graphene and a QW  in a high
magnetic field can be treated as an ideal Bose gas since
magnetoexciton-magnetoexciton interaction vanishes in the limit of
a high magnetic field and a relatively high dielectric constant of the cavity $\epsilon \gg 2$, according to Sec.~\ref{is_mag}. Let us mention that this condition for the high dielectric constant of the microcavity is valid only for graphene, and it is not valid for the quantum well.  Observation of trapped cavity magnetopolaritons
in graphene in a high magnetic field would be an interesting confirmation
of the magnetopolaritonic BEC that we have described. Besides, we have obtained the Rabi splitting related to the creation of a magnetoexciton in a high magnetic field in graphene. Since this Rabi splitting is proportional to $B^{-1/4}$, we conclude that  the Rabi splitting in graphene can be controlled by the external magnetic field $B$, while in a quantum well the Rabi splitting does not depend on the magnetic field when it is strong. The results for the critical BEC temperature of magnetopolaritons in a QW and graphene in high magnetic field are similar, since the magnetoexcitons in both systems are described by the same Hamiltonian.

\acknowledgments
 We would like to thank J.~F. Vazquez-Poritz for the useful discussion.  O.~L.~B., R.~Ya.~K. were supported by PSC CUNY grant 621360040, and Yu.~E.~L. was partially supported by INTAS and RFBR grants.



\end{document}